\documentclass[prd,preprintnumbers]{revtex4}
\usepackage{epsfig}
\usepackage{graphicx}
\usepackage{bm}

\newcommand{\be}{\begin{equation}}
\newcommand{\dd}{\displaystyle}
\newcommand{\ee}{\end{equation}}
\newcommand{\bea}{\begin{eqnarray}}
\newcommand{\eea}{\end{eqnarray}}

\newcommand{\de}{\partial}

\begin{document}
\preprint{{\bf BARI-TH 479/04}} \preprint{{\bf DFF-412/03/04}}
\title{Effective Gap Equation for the Inhomogeneous LOFF
Superconductive Phase}

\author{R. Casalbuoni}
\affiliation{Department of Physics, University of Florence, and
INFN-Florence , Italy}
\author{M. Ciminale, M. Mannarelli,  G. Nardulli, M. Ruggieri}
\affiliation{Department of Physics, University of Bari and
INFN-Bari, Italy}\author{R. Gatto} \affiliation{Department of
Physics, University of Geneva, Switzerland}

\begin{abstract}We present an approximate gap equation for
 different crystalline
structures of the LOFF phase of high density QCD at T=0. This
equation is derived by using an effective condensate term obtained
by averaging the inhomogeneous condensate over distances of the
order of the crystal lattice size. The approximation is expected
to work better far off any second order phase transition. As a
function of the difference of the chemical potentials of the up
and down quarks, $\delta\mu$, we get that the octahedron is
energetically favored from $\delta\mu=\Delta_0/\sqrt 2$ to
$0.95\Delta_0$, where $\Delta_0$ is the gap for the homogeneous
phase, while in the range $0.95\Delta_0-1.32\Delta_0$ the face
centered cube prevails. At $\delta\mu=1.32\Delta_0$ a first order
phase transition to the normal phase occurs.
\end{abstract}
 \maketitle
\section{Introduction \label{sec:0}}

The existence of an inhomogeneous superconductive phase
characterized by a periodic structure of the gap parameter was
hypothesized long ago in two separate and contemporary papers by
Larkin and Ovchinnikov \cite{LO} and by Fulde and Ferrel \cite{FF}
and is therefore named  LOFF phase. These original papers studied
 superconductors in presence of a strong magnetic field
   coupled to the spins of the conduction electrons. The corresponding interaction hamiltonian
    gives rise to a  separation of the Fermi surfaces  of the  pairing electrons since they have
    opposite spins.
    If the separation is very high the pairing is
   destroyed and there is a transition  from the superconductive to the normal state.
   In \cite{LO} and  \cite{FF} it
   was  shown that a new state could be
   formed, close to the transition line. This state has the feature
   of exhibiting an order parameter, the gap, which is not a
   constant, but has a space variation whose typical wavelength is of the
   order of the inverse of the difference in the Fermi energies of the  pairing electrons.
   The space modulation of the gap arises because the electron pair has
   non zero total momentum. In the simplest case the modulation is that of a plane
   wave. In the general case it is  given by the sum
   of several plane waves, which might lead to a
 crystalline structure. In any case the LOFF phase  breaks
   translational and rotational symmetries.

  Inhomogeneous  superconductivity in metals  has been the object of intense
   experimental investigations especially in the last decade, though the evidence is still
   inconclusive (for a review see
   \cite{Casalbuoni:2003wh}). Quite recently
    it has been  also realized that at moderate density
    the mass difference
   between the strange and the up and down quarks at the weak
   equilibrium leads to a difference in their
   Fermi momenta, which renders in principle the  LOFF state
possible in color
   superconductivity \cite{Alford:2000ze} (for general reviews of color
superconductivity see
\cite{Rajagopal:2000wf,Rischke:2000pv,Schafer:2000et,Hong:2000ck,Alford:2001dt,Reddy:2002ri,Nardulli:2002ma}).
 The study of the crystalline
color superconductive phase of QCD is theoretically important for
the understanding of the QCD phase diagram. It may be also
relevant to the astrophysics of compact stars. As a matter of fact
whereas at extreme densities one expects the color-flavor-locking
phase to occur, at lower densities, within the inner core of
compact stars, there is a better chance that the crystalline phase
be realized. In these conditions a difference in the chemical
potentials of the up and down quark might be generated by
beta-equilibrium of the quark matter. While difficult to be
studied in the laboratory this phenomenon might be therefore
relevant in astrophysics, for example by explaining the sudden
   variations (glitches) of the rotation period of the pulsars
   (for a review of color superconductivity
in compact stars see e.g.\cite{Alford:2000sx},
\cite{Alford:2001mh}). In
   \cite{Bowers:2002xr} a Ginzburg-Landau (GL) analysis of several crystalline structures
  was performed for QCD at $T=0$. This approximation holds for
 small values of the gap parameter and therefore this analysis
 studies a small region of the phase diagram, i.e. the corner where the
 difference of the chemical potentials of the pairing quarks $\delta\mu$ is
 near to $\delta\mu_2$, the
 value where a second order phase transition from the
 LOFF to the normal phase is assumed to exist. For the
 one-plane-wave structure $\delta\mu_2$ is not far from the Clogston-Chandrasekhar
 \cite{clogston},
 \cite{chandrasekhar}  point $\delta\mu_1=\Delta_0/\sqrt
 2$ ($\Delta_0$ is the value of the gap for the superconductive homogeneous
 phase). This is approximately the point where a first order transition occurs between the
 homogeneous Bardeen-Cooper-Schrieffer (BCS) and the  inhomogeneous LOFF phases.
   In other cases $\delta\mu_2$ can be larger and the phase transition can
   be first-order.

   An effective field theory for the crystalline color  superconductive
phase  of QCD was developed in \cite{Casalbuoni:2001gt}. The
papers  \cite{Casalbuoni:2002pa} and \cite{Casalbuoni:2002hr}
studied the properties of the Nambu-Goldstone Bosons (NGB)
associated to the breaking of the translation symmetry, whose
properties were analyzed in \cite{Casalbuoni:2002my}, where an
approximation scheme valid at the GL point ($\Delta\to 0$) was
developed. In the present paper we come back to this problem with
the aim to implement an approximation scheme for the gap equation
along similar lines. We will find that an effective gap equation
can be indeed written down, but it is expected to work better for
large $\Delta$, for example close to the BCS/LOFF transition point
$\delta\mu_1$. We will work at $T=0$ and for  high densities,
since we wish to study the LOFF phase in the context of QCD, where
the physically relevant case is presumably the zero temperature
high density inner core of compact stars. We will not consider a
complete analysis, which is of course an endless task, but we
shall limit our analysis to four crystalline structures.  In
Section \ref{sec:1} we discuss the one plane wave as the prototype
case. Then we consider those cases that various analyses
\cite{LO}, \cite{Bowers:2002xr}, \cite{combescot:2003wz} indicate
among the favored forms of the condensate:   the two-antipodal
plane wave, the octahedron and the face-centered-cube, to be
discussed in Sections \ref{sec:3} and \ref{sec:4}. The numerical
results of Section \ref{sec:4} are that, as a function of the
difference $\delta\mu$ between the quark chemical potentials,  the
octahedron is energetically favored from $\delta\mu=\Delta_0/\sqrt
2$ to $0.95\Delta_0$, while in the range
$0.95\Delta_0-1.32\Delta_0$ the face centered cube prevails. After
this point the  normal phase is restored.

\section{The effective gap equation for the Fulde-Ferrel  (one-plane wave) state
\label{sec:1}} We begin our study by a review of one-plane wave,
Fulde-Ferrel (FF) state. Although this case can be solved exactly,
it is useful to consider it here in order to fix the notations and
introduce some definitions to be used later on. We shall consider
Cooper pairing of the massless quarks up $u$ and down $d$, with
chemical potential $\mu_u$, $\mu_d$. We define
$\mu=(\mu_u+\mu_d)/2$ and $\delta\mu=|\mu_u-\mu_d|/2\ll\mu$.

The gap equation, as discussed in Ref. \cite{Casalbuoni:2003wh},
can be derived using the following effective interaction
Lagrangian \be {\cal L}_I=-\frac 3 8 g\bar\psi\gamma^\mu
\lambda^a\psi\,\bar\psi\gamma^\mu
\lambda^a\psi\label{eq:291}\,.\ee Here $g$ is a coupling constant,
$\lambda_a$ are color matrices and a sum over flavors is
understood.

 Following the same steps as in
\cite{Casalbuoni:2003wh}  we get that, in the mean field
approximation, ${\cal L}_I$ in (\ref{eq:291}) is substituted by
\be{\cal L}_{cond}=-\frac 1 2\epsilon_{\alpha\beta
3}\epsilon^{ij}(\psi_i^\alpha\psi_j^\beta\,\Delta(\bm r)+\,{\rm
c.c.})\,+\,(L\to R)-\frac 1 g\Delta(\bm r)\Delta^*(\bm
r)\,,\label{8}\ee where $i,j$ are flavor indices and
$\alpha,\beta$ are color indices. In the FF state the total
momentum of the Cooper pair is $2\bf q$ and the condensate has the
space-dependence of a single plane wave \cite{FF}: \be \Delta(\bm
r)=\Delta e^{2i{\bf q}\cdot{\bf r}}\, .\ee It follows that at zero
temperature, the gap equation for the FF state can be written as
follows\be 1=\frac{g\rho}2\int\frac{d{\bf
v}}{4\pi}\int_0^{\delta}\frac{d\xi}{\sqrt{\xi^2+\Delta^2}}\,
\left(1-\theta(-E_u)-\theta(-E_d)\right)\,,\label{eq:1}\ee where
$|{\bf v}|=1$ and $\delta$ is the ultraviolet cutoff that we
assume equal to $\mu/2$; moreover we assume $\delta\gg q\sim
\delta\mu$. The following equations \be E_{u,d}=\pm\delta\mu\mp
{\bf q} \cdot {\bf v}+\sqrt{\xi^2+\Delta^2}
\label{dispersionFF}\ee provide the quasi-particle dispersion laws
and $\rho$ is the density of states that in QCD with two flavor
quarks is\be \rho=\frac{4\mu^2}{\pi^2}\ .\ee We observe that\be
1-\theta(-E_u)-\theta(-E_d)=\theta(E_u)\theta(E_d)\ee so that the
integration in Eq.(\ref{eq:1}) is over the  pairing region (PR),
defined by \be\label{PRFF} PR = \left\{ (\xi,{\bf v}) \,|\, E_u>0
\,\mathrm{and}\, E_d>0 \right\} \, .\ee More explicitly the
pairing region ($PR$) is defined by the condition (${\bf\hat
q}={\bf q}/q$) \be\text{ Max}\left\{-1,\,
z_q-\frac{\sqrt{\xi^2+\Delta^2}}q\right\}<\,{\bf v\cdot\hat
q}\,<\text{
Min}\left\{1,\,z_q+\frac{\sqrt{\xi^2+\Delta^2}}q\right\}\label{eq:2}\ee
with \be z_q=\frac{\delta\mu}q\ .\ee
 Eq. (\ref{eq:1}) can be
written in a way that will  be useful later: \be
\Delta=i\frac{g\rho}{2}\int\frac{d{\bf v}}{4\pi} \int_0^{\delta}
\frac{d \xi}{2\pi} \int d \ell_0  \,\,
\frac{\Delta_{eff}}{\ell_0^2-\xi^2-\Delta^2_{eff}}\label{GAPeff}
\, , \ee that after energy integration gives \be \Delta=
\frac{g\rho}{2}\int
 \frac{d\,{\bf v}}{4\pi}
 \int_0^\delta\,d\xi\,\frac{\Delta_{eff}}{\sqrt{\xi^2+\Delta^2_{eff}}}
\label{GAP8}\, .\ee Here $\Delta_{eff}=\Delta_{eff}({\bf v\cdot
\hat q}, \xi)$ is defined as\bea
\Delta_{eff}\,=\Delta\theta(E_u)\theta(E_d)\,=\,{\dd\Bigg \{ }
\begin{array}{cc}\Delta & \textrm{~~for~~}
(\xi,{\bf v}) \in PR\cr&\cr 0 &
\textrm{~~elsewhere~~} \ .\\
\end{array}
\label{GAP8.1} \eea

To the interaction  term (\ref{8}) in the lagrangian one should
add the Lagrangian for free quarks. We adopt for its description
the formalism of the High Density Effective Theory (HDET) (see
\cite{Hong:1998tn,Hong:1999ru,Beane:2000ms,Casalbuoni:2000na,Schafer:2003jn},
and, for reviews,
\cite{Casalbuoni:2001dw,Nardulli:2002ma,Schafer:2003yh}) and write
it as follows  \be {\cal L}_{0} = \sum_{\vec v}
\Big[\psi_+^\dagger iV\cdot
\partial\psi_+\ +\
\psi_-^\dagger i\tilde V \cdot \partial\psi_-\Big]\ +\ (L\to
R)\,.\ee Here the sum represents an average over velocities. The
velocity dependent left-handed fields $\psi_\pm\equiv \psi_{\bf\pm
v}$ are the positive energy part in the decomposition
\be\psi(x)=\int\frac{d\bf v}{4\pi} e^{-i\mu v\cdot x}
 \left[\psi_{\bf v}(x)+\Psi_{\bf v}(x)\right]\ \ee while
$\Psi_{\bf v}$ is the negative energy part which has been
integrated out. $\psi_{\bf v}$ is given by \be \psi_{\bf
v}(x)=e^{i\mu v\cdot
x}P_+\psi(x)=\int_{|\ell|<\delta}\frac{d^4\ell}{(2\pi)^4}e^{-i\ell\cdot
x}P_+\psi(\ell)\ee and therefore contains the residual momentum
$\ell$, corresponding to the decomposition of the quark momentum
 $p=\mu v+\ell$, with
$v^\mu=(0,{\bf v})$ and $\ell_\parallel={\bm \ell\cdot}{\bf
v}=\xi$. We also introduce $V^\mu=(1,{\bf v})$ and $\tilde
V^\mu=(1,\,-{\bf v})$. $P_\pm$ are projectors defined by \be
 P_\pm= \frac{1}2 \left(1\pm\gamma_0{{\bm\gamma}\cdot{\bf v}}\right)
  \ .\ee
 In this formalism we get from (\ref{8})  ($C=i\sigma_2$ is the
 charge conjugation matrix)
 \bea
{\cal L }_{cond}&=&-\frac{\Delta} 2\,\sum_{\bf v_i,\bf v_j}
\exp\{i{\bf r}\cdot{\bm\alpha}(\bf v_i,\,\bf v_j,\,{\bf q}
)\}\epsilon_{ij}\epsilon_{\alpha\beta 3}\psi^T_{\bf
v_i;\,i\alpha}(x)C \psi_{-\,\bf v_j;\,j\beta}(x)\cr && -(L\to
R)+{\rm h.c.}-\frac 1 g\Delta(\bm r)\Delta^*(\bm r)\
,\label{loff60}\eea where \be{\bm\alpha}({\bf v_i},\,{\bf
v_j},\,{\bf q})=2{\bf q}-\mu_i{\bf v_i}-\mu_j{\bf v_j}\,.
\label{Lcond}\ee This Lagrangian  can be put, in momentum space,
in the following form \be {\cal L}_{int}=-\frac{1} 2\,
\,\sum_{\vec v}\Delta_{eff}\,\epsilon_{ij}\epsilon_{\alpha\beta
3}\psi^T_{{\bf v};\,i\alpha}(\ell)C \psi_{-\,{\bf
v};\,j\beta}(-\ell)-(L\to R)+{\rm h.c.}-\frac 1 g\Delta\Delta^* \,
,\label{loff60bis}\ee
 by an average procedure (see below) and then we can  obtain the gap equation
(\ref{GAPeff}). It is useful to introduce the following basis for
the fermion fields: \be \psi_{+,\alpha i}=
\sum_{A=0}^3\frac{(\sigma_A)_{\alpha i}}{\sqrt 2}\psi_{+}^A,
~(i,\,\alpha=1,\,2),~~~ \psi_{+,3 1}=\psi_{+}^4,~~~ \psi_{+,3
2}=\psi_{+}^5\ ,\ee where $\sigma_A$ are the Pauli matrices for
$A=1,2,3$ and $\sigma_0=1$ (similar expressions hold for
$\psi^A_-$) and then  we obtain that the propagator for the
velocity dependent fermionic fields is given by \be
D_{AB}(\ell)=\frac{1}{V\cdot \ell\,\tilde V\cdot
\ell\,-\,\Delta_{eff}^2} \left(
\begin{array}{cc}
\tilde V\cdot\ell\,\delta_{AB} &  -\Delta_{AB}
\\
 -\Delta_{AB}
 &
 V\cdot\ell\,\delta_{AB}
\end{array}
\right)\label{propagatore}\ .\ee The  matrix $\Delta_{AB}$ is as
follows: $ \Delta_{AB}=0$ $(A\,{\rm or} \,B=4\,{\rm or} \,5 )$,
and, for $A,B=0,...,3$:\be \Delta_{AB}=\Delta_{eff}
\left(\begin{array}{cccc}
   1& 0 & 0 & 0 \\
  0 & -1 &0 & 0 \\
  0 & 0&-1& 0 \\
  0 & 0 & 0 & -1
\end{array}\right)_{AB}\ .\label{eq:38}
\ee

To obtain the Lagrangian (\ref{loff60bis}) one can perform a
weighted average of the Lagrangian (\ref{loff60}) over a region of
the size of the lattice cell. The weight function $ g_R({\bf r})$
will be defined below. We note that in the gap equation the
relevant momenta are small with respect to the gap which is of the
order of $q$. Therefore we may assume that the velocity dependent
fields are slowly varying over regions of the order of the lattice
size. This means that in the average we can treat them as
constant, and in conclusion the average is made only on the
coefficient $ \exp\{i{\bf r}\cdot{\bm\alpha}\}$. Therefore what we
are computing is \be I ({\bm\alpha})=\Big< \exp\{i{\bf
r}\cdot{\bm\alpha}\}g_R({\bf r})\Big>\ \label{II}\ee where the
bracket means average over the cell.

\begin{figure}[h] \centerline{
\epsfxsize=6cm\epsfbox{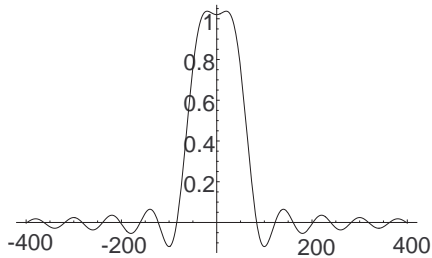} } \caption {{ \rm $I(x)$ as
obtained by Equations (\ref{II}) and (\ref{gII}) (in one
dimension) for $q=40$ MeV and $R=2\pi/3$. \label{deltaerre} }}
\end{figure}

It is possible to choose $ g_R({\bf r})$ in such a way that \be I
({\bm\alpha})= \delta_R^3\left(\frac{\bm\alpha}{2q}\right)\ee
where \be \delta_R^3({\bf x})= {\dd\Bigg \{ }
\begin{array}{cc} {1} & \textrm{~~for~~}
|{\bf x}| < \dd{\frac{\pi}{2R}} \, , \cr&\cr
0 & \textrm{elsewhere}\, . \\
\end{array}
\label{deltaII}\ee An approximate expression of  $ g_R({\bf r})$
is given by
 \be g_R({\bf r})=\prod_{k=1}^3 \frac{\sin\left[\dd\frac{\pi
qr_k}{R}\right]}{\pi r_k}\,. \label{gII}\ee For $R/\pi\approx 1$,
$g_R$ is different from zero only in a region of the size of the
unit lattice cell; the resulting integral $I$ obtained by
(\ref{II}) and (\ref{gII})
 is reported (for the  one-dimensional case and for
$R=2\pi/3$) in fig.\ref{deltaerre}.

Independently of the exact form of $ g_R({\bf r})$, we will assume
that the average procedure gives as a result the brick-shaped
function $\delta_R$ defined in (\ref{deltaII}). Clearly this
assumption is a crucial ingredient of our approximation.

 In the $\mu\to\infty$ limit by taking $\bf q$ along
the $z-$axis, we get for the components $x$ and $y$ of $\bm\alpha$
\be |(\mu_1 v_1+\mu_2 v_2)_{x,y}|<\frac{\pi q}R\,,\ee i.e.
approximately (for $\delta\mu\ll\mu$) \be |( v_1+
v_2)_{x,y}|<\frac{\pi q}{R\mu}\,.\label{25b}\ee From this in the
high density limit it follows \be{\bf v_1}= -{\bf v_2}+{\cal
O}(1/\mu)\,.\ee We use the $x$ and $y$ components of the
$\delta_R$ function to get rid of the integration over ${\bf v}_2$
in eq. (\ref{loff60}). The two factors $\pi/R$ arising from the
$x$ and $y$ components are absorbed into a wave function
renormalization of the quark fields, both in the kinetic and in
the gap terms. The $z$ component gives a factor $\delta_R[h({\bf
v\cdot\hat q})]$, where the function $h$ was computed in
\cite{Casalbuoni:2002my}. Since here we work in the limit
$\delta\mu\ll\mu$, we neglect corrections of order $\delta\mu/\mu$
and we use the asymptotic expression \cite{Casalbuoni:2002my}:\be
h({\bf v\cdot\hat q})=1-\frac{z_q}{\bf v\cdot\hat
q}\,.\label{hdix}\ee After this average we get that the
coefficient of the bilinear term in Eq.(\ref{Lcond}) is \be
\Delta\,\delta_R[h({\bf v\cdot\hat q})]\ .\ee This factor
coincides with $\Delta_{eff}({\bf v\cdot\hat q},\xi)$ if one
chooses \be R=\frac{\pi|\delta\mu-{\bf v\cdot\bf
q}|}{2\sqrt{\xi^2+\Delta^2}|h({\bf v\cdot\hat q})|}\label{eq:31}
\, . \ee
 This is consistent with our hypothesis, $R/\pi\approx 1$, only  if
 $\Delta$ is not too small, meaning that we should be far from a
 second order phase transition.

 Let us finally notice that  Eq. (\ref{GAP8}) can
 be obtained from Eq. (\ref{GAPeff}) by substituting the previous
 expression for $R$ with
\be R=\frac{\pi|\delta\mu-{\bf v\cdot\bf
q}|}{2|\ell_0|\cdot|h({\bf v\cdot\hat q})|}\ .\label{R}\ee In
fact, observing that in any case $\Delta_{eff}$ is equal to 0 or
$\Delta$, at the pole we get back the expression (\ref{eq:31}).
Then written as in  (\ref{R}), $R$ and analogously $\Delta_{eff}$
become  functions of the velocity and the energy; therefore our
average should be better taken in the momentum space.

\section{The effective gap equation at $T=0$ for generic crystalline
structures\label{sec:3}} Let us now consider the general case of
$P$ plane waves\be \Delta(\bm r)=\sum_{m=1}^P\Delta_m e^{2i{\bf
q_m}\cdot{\bf r}}\label{de}\ .\ee We shall consider only the case
$\Delta_m=\Delta$ for any $m=1,...P$ and \be{\bf q_m}\,
 =\,q\,{\bf n_m}\, ,\label{qn}\ee
 with  ${\bf n_m}$  unit vectors.
Generalizing the results of the previous equations we substitute
in the Lagrangian (\ref{loff60bis}) and in the propagator
(\ref{propagatore})  $\Delta_{eff}\left({\bf v\cdot {\bf
n}},\ell_0\right)$ with \be\Delta_{E}({\bf
v},\ell_0)=\sum_{m=1}^P\Delta_{eff}\left({\bf v\cdot\bf
n_m},\ell_0\right)\label{eq13}\, .\ee  In this equation we have
made explicit the dependence on the energy $\ell_0$ arising from
the average. Correspondingly the gap equation is \be P
\Delta=i\frac{g\rho}{2}\int\frac{d{\bf v}}{4\pi}\int
\frac{d^2\ell}{2\pi} \, \frac{\Delta_{E}({\bf
v},\ell_0)}{\ell_0^2-\ell_\parallel^2-\Delta_{E}^2({\bf
v},\ell_0)}\label{29} \ \ee which generalizes Eq. (\ref{GAPeff}).
The origin of the factor $P$ on the l.h.s. of this equation is as
follows. The Lagrangian contains the term \be\frac{\Delta^*({\bf
r})\Delta({\bf
 r})}g
\ee which, when averaged over the cell, gets non vanishing
contribution only from the diagonal terms in the double sum over
the plane waves.

 The energy integration is performed by the residue theorem and
 the phase space is divided into different regions according to
 the pole positions, defined by \be \epsilon=\sqrt{\xi^2+\Delta^2_{E}({\bf
 v},\epsilon)}\ .\ee Therefore we get \be
 P
\Delta\ln\frac{2\delta}{\Delta_0} =\sum_{k=1}^P \int\int_{P_k}
\frac{d{\bf v}}{4\pi} d\xi \, \frac{\Delta_{E}({\bf
v},\epsilon)}{\sqrt{\xi^2+\Delta^2_{E}({\bf v},\epsilon)}}
 =\sum_{k=1}^P\int\int_{P_k}\frac{d{\bf v}}{4\pi} d\xi \,
\frac{k\Delta}{\sqrt{\xi^2+k^2\Delta^2}}\label{290} \ \ee where
the regions $P_k$ are defined as follows \be P_k=\{({\bf
v},\xi)\,|\,\Delta_E({\bf v},\epsilon)=k\Delta\}\ \ee and
 we have made use of the equation\be \frac 2{g\rho} =
\ln\frac{2\delta}{\Delta_0}\ee relating the BCS gap $\Delta_0$ to
the four fermion coupling $g$ and the density of states.  The
first term in the sum, corresponding to the region $P_1$, has $P$
equal contributions with a dispersion rule equal to the Fulde and
Ferrel case. This can be interpreted as a contribution from $P$
non interacting plane waves. In the other regions the different
plane waves have an overlap. Since the definition of the regions
$P_k$ depends on the value of $\Delta$, their determination is
part of the problem of solving the gap equation.

Let us comment on our result. We have shown that in our
approximation the dispersion relation for the quasi-particles has
several branches corresponding to the values $k\Delta$,
$k=1,\cdots,P$. Therefore the following  interpretation of the gap
equation (\ref{290}) can be given. Each term in the sum
corresponds to one branch of the dispersion law, i.e. to the
propagation of a gapped quasi-particle with gap $k\Delta$. The
corresponding region is nothing but $P_k$. However, the regions
$P_k$ do not represent a partition of the phase space since it is
possible to have at the same point quasi-particles with different
gaps.

As a final remark, it is possible to test our method in an exactly
solvable case, i.e.  the particular case of $P$ identical plane
waves. In this case  the only non empty region in the gap equation
is $P_P$ therefore the gap equation reduces to the FF gap equation
for a gap $P\Delta$, i.e. the expected result. We assume this as a
 consistency test of our approximation.

\section{Numerical analysis\label{sec:4}}
The free energy $\Omega$ is obtained by integrating in $\Delta$
the gap equation.
 At fixed $\delta\mu$, $\Omega
$ is a function of $\Delta$ and $z_q$, therefore the energetically
favored state satisfies the conditions
 \be \frac{\de \Omega}{\de \Delta} \, =\, 0 \, ,\hspace{1.cm}
\frac{\de \Omega}{\de z_q} \, =\, 0 \,,\ee and must be the
absolute minimum. We now analyze three different structures by
this criterion. We will assume $\mu=400$~MeV and $\delta=\mu/2$.
\subsection{Strip}

This is the case of two antipodal plane waves, i.e. we take $P=2$
in Eq.(\ref{de}) and \be {\bf n}_1=(0,0,+1)\,,\hskip1cm {\bf
n}_2=(0,0,-1)\,,\ee in Eq.(\ref{qn}). If $\epsilon$ is the pole
position in the $\ell_0$ complex plane, the gap equation is \be
 2
\Delta\ln\frac{2\delta}{\Delta_0} = \int\int_{P_1} \frac{d{\bf
v}}{4\pi} d\xi \, \frac{\Delta_{E}({\bf
v},\epsilon)}{\sqrt{\xi^2+\Delta^2_{E}({\bf v},\epsilon)}} +
\int\int_{P_2} \frac{d{\bf v}}{4\pi}d\xi \, \frac{\Delta_{E}({\bf
v},\epsilon)}{\sqrt{\xi^2+\Delta^2_{E}({\bf v},\epsilon)}}
\label{46} \, .\ee

Here $\dd \int_{P_1}$ represents a region where the two
$\Delta_{eff}$ appearing in $\Delta_E$ have no overlap. In this
region \be\Delta_{E}({\bf v},\epsilon)=\Delta\left\{ \theta
(E^1_u)\theta (E^1_d)+\theta (E^2_u)\theta (E^2_d)-2 \theta
(E^1_u)\theta (E^1_d)\theta (E^2_u)\theta
(E^2_d)\right\}_{\Delta}=\Delta\, . \ee In other words \be
\int\int_{P_1} \frac{d{\bf v}}{4\pi} d\xi \, \frac{\Delta_{E}({\bf
v},\epsilon)}{\sqrt{\xi^2+\Delta^2_{E}({\bf v},\epsilon)}} =\Delta
\int\int \frac{d{\bf v}}{4\pi} d\xi \, \frac{ \left\{ \theta
(E^1_u)\theta (E^1_d)+\theta (E^2_u)\theta (E^2_d)-2 \theta
(E^1_u)\theta (E^1_d)\theta (E^2_u)\theta (E^2_d)
\right\}_{\Delta}}{\sqrt{\xi^2+\Delta^2}} \, . \ee
 On the other hand the integral over $P_2$
 represents the region where the two
$\Delta_{eff}$ appearing in $\Delta_E$ do overlap. In this region
$\Delta^2_{E}({\bf v},\epsilon)=4\Delta^2$ and we have

\be \int\int_{P_2} \frac{d{\bf v}}{4\pi} d\xi \,
\frac{\Delta_{E}({\bf v},\epsilon)}{\sqrt{\xi^2+\Delta^2_{E}({\bf
v},\epsilon)}} =\,\Delta \int\int \frac{d{\bf v}}{4\pi} d\xi \,
\frac{2\, \left\{\theta (E^1_u)\theta ( E^1_d)\theta (E^2_u)\theta
(E^2_d)\right\}_{2\Delta} }{\sqrt{\xi^2+4\Delta^2}}
 \, ,\ee where the subscript $2 \Delta$ on the r.h.s. means that in the
 dispersion laws $E_{u,d}^{1,2}$ one has to use Eqns.
 (\ref{dispersionFF}) with $\Delta\to 2\Delta$.
As already observed, a test of our approximation is given by the
fact that the gap equation obtained in this way reproduces in
 the case of ${\bf q}_1={\bf q}_2$ the FF result.

As we observed above, we expect more reliable results for large
$\Delta$, i.e. far off the GL point. Nevertheless we can determine
the phase transition point $\delta\mu_2$ within this
approximation; we find $\delta\mu_2\approx 0.83\Delta_0$, which
compares fairly well with the exact result in the weak coupling
limit, i.e. $0.75\Delta_0 $; in our approximation we find a first
order phase transition (see subsection \ref{sub} for further
comments).

\begin{figure}[h] \centerline{
\epsfxsize=5cm\epsfbox{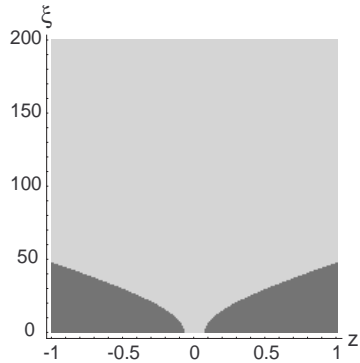} } \caption {{ \rm
The regions $P_1$ and $P_2$ for the strip in the $(z,\xi)$ plane
with $z=\cos\theta= {\bf v \cdot \hat q_1}$. The region $P_1$ is
the dark region, whereas the region $P_2$ covers all the plane and
is in light grey. Here $\Delta= 0.75 \Delta_0$, $z_q=1.0 $,
$\delta\mu = \delta\mu_1$; $\xi$ in MeV.\label{pairing-striscia1}
}}
\end{figure}

It may be useful to have a pictorial representation of the various
regions. They are reported for the strip in a $(z=\cos\theta,\xi)$
plane in Fig. \ref{pairing-striscia1}, at $\delta\mu=\delta\mu_1$,
where our approximation is expected to work better.

To get a better insight of the main features of our approximation,
we can compare the approximate dispersion law we are using with
the exact one.  The exact dispersion law for the strip was derived
in \cite{LO} (see also \cite{Casalbuoni:2003sa}; in both papers
the energy is computed for each species  from the corresponding
Fermi surface):

\be E^1_{u,d}=\pm\delta\mu\mp {\bf q} \cdot {\bf v}+w|\xi|,~~~
E^2_{u,d}=\pm\delta\mu\pm {\bf q} \cdot {\bf v}+w|\xi|
\label{exact}\ee where \be
w^{-1}=I_0\left(\frac{2\Delta}{q\cos\theta}\right)\ .\ee $I_0$ is
the Bessel function and $\theta$ the angle between the $z-$axis
and ${\bf q}={\bf q_1}$.

\begin{figure}[h] \centerline{
\epsfxsize=10cm\epsfbox{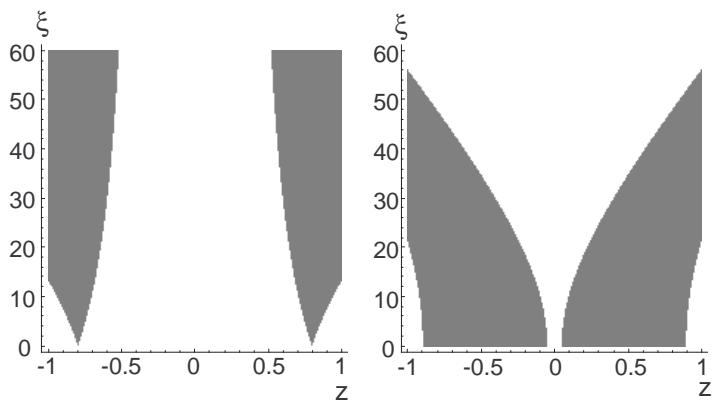} } \caption {{ \rm On the
left: Pairing (in gray) and blocking (in white) regions as
computed by the exact dispersion law (\ref{exact}) for the strip.
On the right: The gray region depicts the area where coupling is
more favourable (here both quasiparticles with gap $\Delta$ and
$2\Delta$ contribute); in the white area only quasiparticles with
gap $2\Delta$ can be  present. On the horizontal axis
$z=\cos\theta$; on the vertical axis the longitudinal momentum
measured from the Fermi surface, $\xi$ (in MeV). Here $\Delta=
0.75 \Delta_0 $ $z_q=0.8 $, $\delta\mu = \delta\mu_1$.
\label{confronto2} }}
\end{figure}

 This comparison for the strip is reported in Fig. \ref{confronto2}
(we use the parameters $\Delta= 0.75 \Delta_0 $, $\delta\mu =
\delta\mu_1$ and $z_q\approx 0.8 $, which is more appropriate for
comparison with the exact result). On the right diagram the gray
region depicts the area where coupling is more favorable, because
here both quasiparticles with gap $\Delta$ and $2\Delta$
contribute; to this area corresponds on the left diagram, in gray,
the pairing region, where $ E^k_{u,d}>0$ ($k=1,2$). The white
areas depict respectively, on the right an area where coupling is
less favorable, because only quasiparticles with gap
 $2\Delta$ can contribute, and on the left the blocking region, where
 the energies are negative. We have limited the analysis to the
 region of smaller $\xi$ since this is the region more relevant
 for coupling. The two graphs do not coincide, which is due to
 the approximation, however they share some features, e.g.
 for both the central region ($z\approx 0$) is disfavored; also
 the areas with small $\xi$ and large $|\cos\theta|$ are less
 favorable in both graphs.
 This leads to the following interpretation of the different regions of the phase space.
 The region where only the quasiparticles of gap $2\Delta$ are
 present corresponds {\it grosso modo} to the blocking region of the exact
 dispersion law, whereas the region where both species of
 quasiparticles (i.e. with gaps $\Delta$ and $2\Delta$) are
 present corresponds to the pairing region of the exact dispersion
 law. This interpretation is supported by the numerical results
 for the gap equation. In fact we find that the contribution of the former
 region (corresponding to blocking) to the gap equation is only 20\% of the total.

\subsection{Octahedron and Face
Centered Cube} For the octahedron (i.e. body-centered-cube, bcc)
we have $P=6$ in (\ref{de}) and we take \bea &&
 {\bf   n_1}=(+1,0,0),~~~ {\bf n_2}=(0,+1,0),~~~{\bf n_3}=(0,0,+1),\cr &&
{\bf n_4}=(-1,0,0), ~~~{\bf n_5}=(0,-1,0),~~~ {\bf n_6}=(0,0,-1)~
\label{eq:168}\eea in (\ref{qn}). For the face centered cube (fcc)
we have $P=8$ and \bea &&
 {\bf   n_1}=\frac{1}{\sqrt 3}(+1,+1,+1),~~~ {\bf n_2}=\frac{1}{\sqrt
3}(+1,-1,+1),~\cr && {\bf n_3}=\frac{1}{\sqrt 3}(-1,-1,+1),~~~
{\bf n_4}=\frac{1}{\sqrt 3}(-1,+1,+1),\cr && {\bf
n_5}=\frac{1}{\sqrt 3}(+1+,1,-1),~~~ {\bf n_6}=\frac{1}{\sqrt
3}(+1,-1,-1), \cr && {\bf n_7}=\frac{1}{\sqrt 3}(-1,-1,-1),~~~
{\bf n_8}=\frac{1}{\sqrt 3}(-1,+1,-1)~ .\label{eq:169}\eea The
topological structure of the different regions $P_k$ is more
involved for the cubes than for the strip.
\begin{figure}[h] \centerline{
\epsfxsize=12cm\epsfbox{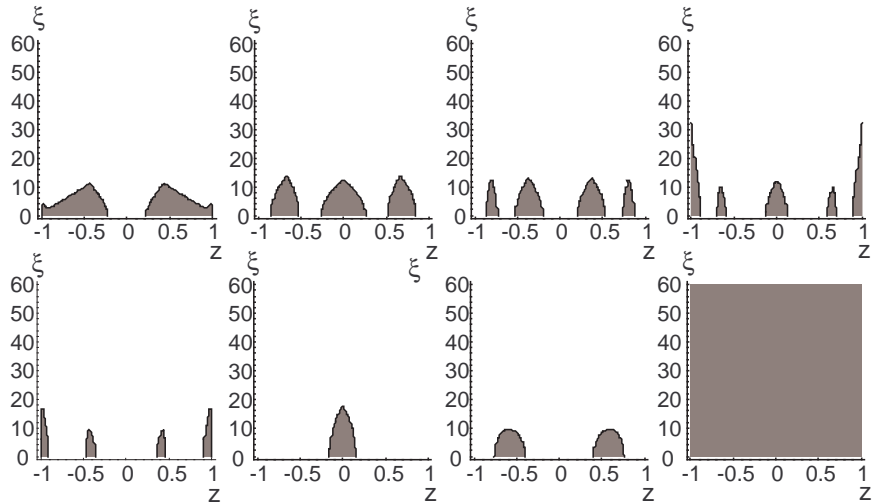}} \caption { \rm   The
pairing regions $P_1,\dots,P_8$ (from left to right and from top
to bottom) for the face centered cube in the $(z,\xi)$ plane for a
fixed value of the polar angle $\varphi = .5$ rad. In each figure
the grey area corresponds to the  pairing region. Here $\delta\mu
= \delta\mu_1$, $\Delta= 0.21 \Delta_0
  $,  $z_q=0.9$. Note that for this particular value of the parameters the $P_8$ region covers all the
  Fermi surface.\label{pairing-fcc}}
\end{figure}
  We simply offer in fig. \ref{pairing-fcc} a picture of
the different pairing regions $P_1,\dots,P_8$  for the face
centered cube in the $(z,\xi)$ plane  at fixed polar angle
$\varphi = 0.5$ rad. In each figure the grey areas correspond to
the  pairing region.

 \subsection{Numerical results\label{sub}}We now present numerical results for
  four crystalline structures (FF, strip, bcc, fcc)
  that are  obtained solving Eq.(\ref{29}) with the appropriate set of
 unit vectors $\bf n_m$.    In  Table \ref{tabledeltamu1} we
 report the
 results obtained at the Clogston point ($\delta\mu=\delta\mu_1$).
Here $P$ is the number of plane waves of each structure; the table
shows that, among the four considered cases, the favored structure
is the octahedron ($P=6$).
\begin{table}[htb]\begin{center}
 \begin{minipage}{6.5in}\begin{tabular}{|c|c|c|c|}
   \hline
  $ P $&$ z_q $& $\frac \Delta{ \Delta_0}
  $&$\frac{2\Omega}{\rho\Delta_0^2}$
\\   \hline
 1& 0.78 & 0.24 & $-1.8\times 10^{-3}$ \\
 2& 1.0 & 0.75 & -0.08 \\
  6& 0.9 & 0.28 & -0.11 \\
  8&0.9& 0.21& -0.09\\ \hline
 \end{tabular}\end{minipage}
\end{center} \caption{\label{tabledeltamu1} The
  gap, $z_q=\delta\mu /q$ and the free energy at  $\delta\mu = \delta\mu_1=\Delta_0/\sqrt 2$
  for different crystalline structures.}\end{table}

\begin{figure}[h] \centerline{
\epsfxsize=5.5cm\epsfbox{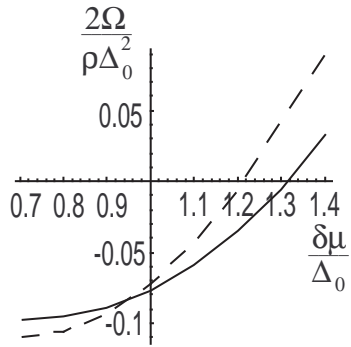} } \caption {{ \rm The values of
the free energies of the octahedron (dashed line) and of the fcc
(full line) crystalline LOFF  structures as a function of
$\delta\mu/\Delta_0$. The octahedron is the favored structure up
to $\delta\mu \approx  .95 \Delta_0$; for $.95 \Delta_0< \delta\mu
< 1.32 \Delta_0$ the fcc is  favored. Here, for each value of
$\delta\mu$, the values of $z_q$ and $\Delta$ are those that
minimize the free energy. \label{omegaplot} }}
\end{figure}

We have analyzed also the case $\delta\mu\neq\delta\mu_1$. In fig.
\ref{omegaplot} we report the plot of the free energies of the
octahedron (dashed line) and of the fcc (full line)  phases as a
function of $\delta\mu/\Delta_0$.

We find that the octahedron is the favored structure up to
$\delta\mu\approx .95 \Delta_0$.  For larger values of
$\delta\mu<1.32\Delta_0$ the favored structure is the fcc. In
Table \ref{tabledeltamu2} we report numerical results for
$\delta\mu_2$ for each crystalline structure and the computed
order of the phase transition between the LOFF and the
 normal phases. We have reported also   the values
of $z_q$ and of the discontinuity in $\Delta/\Delta_0$ at
$\delta\mu=\delta\mu_2-0^+$. The values of $z_q$ and $\Delta$ are
determined minimizing the free energy.
\begin{table}[htb]\begin{center}
 \begin{minipage}{6.5in}
 \begin{tabular}{|c|c|c|c|c|}\hline
$P $ &  $\delta\mu_2/ \Delta_0$  &  Order & $   z_q$ & $ \Delta/
\Delta_0$
\\
\hline
 1& 0.754 &  II&0.83 & 0  \\
 2& 0.83 &I& 1.0 & 0.81  \\
  6& 1.22 & I&0.95 & 0.43 \\
  8& 1.32 &I&0.9& 0.35  \\ \hline
\end{tabular} \end{minipage}
\end{center} \caption{\label{tabledeltamu2} The
values of $\delta\mu_2$, $z_q=\delta\mu /q$, the discontinuity of
$\Delta/\Delta_0$
 and the order of the phase transition between the LOFF and the
 normal phases
  for different crystalline structures.}\end{table}

 As remarked above,  for the strip, in our
approximation neither the order of the transition nor the point
where the transition occurs coincide with those obtained within
the Ginzburg-Landau approximation \cite{LO}. The difference in
$\delta\mu_2$ is $\sim 10\%$ and in $z_q$ is $\sim 17\% $ (the
value for $z_q$ is $0.83$).

\section{Discussion and conclusions\label{sec:5}}

In this paper we have discussed the gap equation for different
Fulde-Ferrel-Larkin-Ovchinnikov phases at $T=0$. Generally
speaking this is a quite complicated problem and the only soluble
case corresponds to the original Fulde-Ferrel phase with a single
plane wave. Typically the discussion of the gap equation is made
via the Ginzburg-Landau approximation which, however, holds only
in proximity of a second order transition. However, the analysis
of Ref. \cite{Bowers:2002xr} shows that in many interesting cases
there is no second order transition to the normal phase.
Therefore, in this paper, we have developed a new kind of
approximation to overcome the problems of the Ginzburg-Landau
treatment. In fact, our approximation holds in the opposite case,
that is when the gap is not too small.  The main feature of our
scheme is a convenient average over the sites defined by the
crystalline structure of the condensate. The description one
obtains amounts to the definition of a multi-valued gap function
having $P$ branches. Each of the branches has a gap given by
$k\Delta$, $k=1,\cdots,P$, with $\Delta$ the solution of the gap
equation. Furthermore each non vanishing value of the gap defines
a particular region in the phase space, the pairing region $P_k$.
It is interesting to notice that each region $P_k$  depends on the
actual value of $\Delta$ and therefore its very definition depends
on the solution of the gap equation itself.

We have noticed that the regions $P_k$ do not represent a
partition of the phase space since it is possible to have at the
same point quasi-particles with different gaps. This observation
would allow to formulate the problem in terms of an effective
Lagrangian containing  $P$ velocity-dependent fields $\psi^{(k)}$,
each of them defined in the region $P_k$, and therefore with gap
$k\Delta$. We will come back to this effective Lagrangian in
future work.

The gap equation has been solved numerically. We have first
analyzed the case $P=1$ where our approximation is exact, and then
the cases $P=2$ (two antipodal plane waves, or strip), $P=6$ (the
octahedron or body-centered-cube) and $P=8$  (the
face-centered-cube). Our main result is that, at $T=0$,  and up to
$\delta\mu \approx 0.95\Delta_0$ the energetically favored
crystalline structure is the octahedron ($P=6$). For larger values
of $\delta\mu/\Delta_0$ the fcc becomes the most favored structure
up to the transition to the normal state occurring at
$\delta\mu_2=1.32\Delta_0$.

Also, in all the cases $P=2,6,8$, we have found that the
transition to the normal state is a first order one.

It is difficult to compare our results to previous work since we
have studied a region of the phase diagram not accessible to the
Ginzburg-Landau approximation. In any event let us summmarize some
of the previous results. Within the GL approximation, the authors
of Ref. \cite{LO} found that the strip was favored in comparison
to the one-plane-wave.  Also in Ref. \cite{Bowers:2002xr} a GL
analysis is performed. These authors study a much larger number of
crystalline structures and conjecture that the most stable case is
the face-centered cube; however no second order transition is
observed for this structure and this makes questionable the
Ginzburg-Landau approximation. As for the octahedron, the phase
transition between the LOFF and the normal phase is also
first-order. These results on the order of the transition for
$P=6,8$  agree with our approximation. Finally, in a recent paper
\cite{combescot:2003wz}, the analogous problem has been studied in
the context of the quasiclassical equations for superconductors by
using a Fourier expansion. These authors get that the favored
state at zero temperature is the octahedron, $P=6$, and also find
that the transition to the normal state is first order. Therefore
the results of their analysis agree with our approximation, at
least qualitatively.


\begin{thebibliography}{29}
\expandafter\ifx\csname
natexlab\endcsname\relax\def\natexlab#1{#1}\fi
\expandafter\ifx\csname bibnamefont\endcsname\relax
  \def\bibnamefont#1{#1}\fi
\expandafter\ifx\csname bibfnamefont\endcsname\relax
  \def\bibfnamefont#1{#1}\fi
\expandafter\ifx\csname citenamefont\endcsname\relax
  \def\citenamefont#1{#1}\fi
\expandafter\ifx\csname url\endcsname\relax
  \def\url#1{\texttt{#1}}\fi
\expandafter\ifx\csname
urlprefix\endcsname\relax\def\urlprefix{URL }\fi
\providecommand{\bibinfo}[2]{#2}
\providecommand{\eprint}[2][]{\url{#2}}

\bibitem[{\citenamefont{Larkin and Ovchinnikov}(1964)}]{LO}
\bibinfo{author}{\bibfnamefont{A.~J.} \bibnamefont{Larkin}} \bibnamefont{and}
  \bibinfo{author}{\bibfnamefont{Y.~N.} \bibnamefont{Ovchinnikov}},
  \bibinfo{journal}{Zh. Exsp. teor. Fiz.} \textbf{\bibinfo{volume}{47}},
  \bibinfo{pages}{1136} (\bibinfo{year}{1964}).

\bibitem[{\citenamefont{Fulde and Ferrell}(1964)}]{FF}
\bibinfo{author}{\bibfnamefont{P.}~\bibnamefont{Fulde}} \bibnamefont{and}
  \bibinfo{author}{\bibfnamefont{R.~A.} \bibnamefont{Ferrell}},
  \bibinfo{journal}{Phys. Rev.} \textbf{\bibinfo{volume}{135}},
  \bibinfo{pages}{A550} (\bibinfo{year}{1964}).

\bibitem[{\citenamefont{Casalbuoni and Nardulli}(2004)}]{Casalbuoni:2003wh}
\bibinfo{author}{\bibfnamefont{R.}~\bibnamefont{Casalbuoni}} \bibnamefont{and}
  \bibinfo{author}{\bibfnamefont{G.}~\bibnamefont{Nardulli}},
  \bibinfo{journal}{Rev. Mod. Phys.} \textbf{\bibinfo{volume}{76}},
  \bibinfo{pages}{263} (\bibinfo{year}{2004}), \eprint{hep-ph/0305069}.

\bibitem[{\citenamefont{Alford et~al.}(2001{\natexlab{a}})\citenamefont{Alford,
  Bowers, and Rajagopal}}]{Alford:2000ze}
\bibinfo{author}{\bibfnamefont{M.~G.} \bibnamefont{Alford}},
  \bibinfo{author}{\bibfnamefont{J.~A.} \bibnamefont{Bowers}},
  \bibnamefont{and}
  \bibinfo{author}{\bibfnamefont{K.}~\bibnamefont{Rajagopal}},
  \bibinfo{journal}{Phys. Rev.} \textbf{\bibinfo{volume}{D63}},
  \bibinfo{pages}{074016} (\bibinfo{year}{2001}{\natexlab{a}}),
  \eprint{hep-ph/0008208}.

\bibitem[{\citenamefont{Rajagopal and Wilczek}(2001)}]{Rajagopal:2000wf}
\bibinfo{author}{\bibfnamefont{K.}~\bibnamefont{Rajagopal}} \bibnamefont{and}
  \bibinfo{author}{\bibfnamefont{F.}~\bibnamefont{Wilczek}},
  \emph{\bibinfo{title}{{\rm in} At the frontier of particle physics, vol. 3}}
  (\bibinfo{publisher}{World Scientific}, \bibinfo{year}{2001}),
  \eprint{hep-ph/0011333}.

\bibitem[{\citenamefont{Rischke and Pisarski}(2000)}]{Rischke:2000pv}
\bibinfo{author}{\bibfnamefont{D.~H.} \bibnamefont{Rischke}} \bibnamefont{and}
  \bibinfo{author}{\bibfnamefont{R.~D.} \bibnamefont{Pisarski}}
  (\bibinfo{year}{2000}), \eprint{nucl-th/0004016}.

\bibitem[{\citenamefont{Schafer and Shuryak}(2001)}]{Schafer:2000et}
\bibinfo{author}{\bibfnamefont{T.}~\bibnamefont{Schafer}} \bibnamefont{and}
  \bibinfo{author}{\bibfnamefont{E.~V.} \bibnamefont{Shuryak}},
  \bibinfo{journal}{Lect. Notes Phys.} \textbf{\bibinfo{volume}{578}},
  \bibinfo{pages}{203} (\bibinfo{year}{2001}), \eprint{nucl-th/0010049}.

\bibitem[{\citenamefont{Hong}(2001)}]{Hong:2000ck}
\bibinfo{author}{\bibfnamefont{D.~K.} \bibnamefont{Hong}},
  \bibinfo{journal}{Acta Phys. Polon.} \textbf{\bibinfo{volume}{B32}},
  \bibinfo{pages}{1253} (\bibinfo{year}{2001}), \eprint{hep-ph/0101025}.

\bibitem[{\citenamefont{Alford}(2001)}]{Alford:2001dt}
\bibinfo{author}{\bibfnamefont{M.~G.} \bibnamefont{Alford}},
  \bibinfo{journal}{Ann. Rev. Nucl. Part. Sci.} \textbf{\bibinfo{volume}{51}},
  \bibinfo{pages}{131} (\bibinfo{year}{2001}), \eprint{hep-ph/0102047}.

\bibitem[{\citenamefont{Reddy}(2002)}]{Reddy:2002ri}
\bibinfo{author}{\bibfnamefont{S.}~\bibnamefont{Reddy}}, \bibinfo{journal}{Acta
  Phys. Polon.} \textbf{\bibinfo{volume}{B33}}, \bibinfo{pages}{4101}
  (\bibinfo{year}{2002}), \eprint{nucl-th/0211045}.

\bibitem[{\citenamefont{Nardulli}(2002)}]{Nardulli:2002ma}
\bibinfo{author}{\bibfnamefont{G.}~\bibnamefont{Nardulli}},
  \bibinfo{journal}{Riv. Nuovo Cim.} \textbf{\bibinfo{volume}{25N3}},
  \bibinfo{pages}{1} (\bibinfo{year}{2002}), \eprint{hep-ph/0202037}.

\bibitem[{\citenamefont{Alford et~al.}(2001{\natexlab{b}})\citenamefont{Alford,
  Bowers, and Rajagopal}}]{Alford:2000sx}
\bibinfo{author}{\bibfnamefont{M.~G.} \bibnamefont{Alford}},
  \bibinfo{author}{\bibfnamefont{J.~A.} \bibnamefont{Bowers}},
  \bibnamefont{and}
  \bibinfo{author}{\bibfnamefont{K.}~\bibnamefont{Rajagopal}},
  \bibinfo{journal}{J. Phys.} \textbf{\bibinfo{volume}{G27}},
  \bibinfo{pages}{541} (\bibinfo{year}{2001}{\natexlab{b}}),
  \eprint{hep-ph/0009357}.

\bibitem[{\citenamefont{Alford}(2002)}]{Alford:2001mh}
\bibinfo{author}{\bibfnamefont{M.}~\bibnamefont{Alford}},
  \bibinfo{journal}{eConf} \textbf{\bibinfo{volume}{C010815}},
  \bibinfo{pages}{137} (\bibinfo{year}{2002}), \eprint{hep-ph/0110150}.

\bibitem[{\citenamefont{Bowers and Rajagopal}(2002)}]{Bowers:2002xr}
\bibinfo{author}{\bibfnamefont{J.~A.} \bibnamefont{Bowers}} \bibnamefont{and}
  \bibinfo{author}{\bibfnamefont{K.}~\bibnamefont{Rajagopal}},
  \bibinfo{journal}{Phys. Rev.} \textbf{\bibinfo{volume}{D66}},
  \bibinfo{pages}{065002} (\bibinfo{year}{2002}), \eprint{hep-ph/0204079}.

\bibitem[{\citenamefont{Clogston}(1962)}]{clogston}
\bibinfo{author}{\bibfnamefont{A.~M.} \bibnamefont{Clogston}},
  \bibinfo{journal}{Phys. Rev. Lett.} \textbf{\bibinfo{volume}{9}},
  \bibinfo{pages}{266} (\bibinfo{year}{1962}).

\bibitem[{\citenamefont{Chandrasekhar}(1962)}]{chandrasekhar}
\bibinfo{author}{\bibfnamefont{B.~S.} \bibnamefont{Chandrasekhar}},
  \bibinfo{journal}{App. Phys. Lett.} \textbf{\bibinfo{volume}{1}},
  \bibinfo{pages}{7} (\bibinfo{year}{1962}).

\bibitem[{\citenamefont{Casalbuoni
  et~al.}(2001{\natexlab{a}})\citenamefont{Casalbuoni, Gatto, Mannarelli, and
  Nardulli}}]{Casalbuoni:2001gt}
\bibinfo{author}{\bibfnamefont{R.}~\bibnamefont{Casalbuoni}},
  \bibinfo{author}{\bibfnamefont{R.}~\bibnamefont{Gatto}},
  \bibinfo{author}{\bibfnamefont{M.}~\bibnamefont{Mannarelli}},
  \bibnamefont{and} \bibinfo{author}{\bibfnamefont{G.}~\bibnamefont{Nardulli}},
  \bibinfo{journal}{Phys. Lett.} \textbf{\bibinfo{volume}{B511}},
  \bibinfo{pages}{218} (\bibinfo{year}{2001}{\natexlab{a}}),
  \eprint{hep-ph/0101326}.

\bibitem[{\citenamefont{Casalbuoni
  et~al.}(2002{\natexlab{a}})\citenamefont{Casalbuoni, Gatto, Mannarelli, and
  Nardulli}}]{Casalbuoni:2002pa}
\bibinfo{author}{\bibfnamefont{R.}~\bibnamefont{Casalbuoni}},
  \bibinfo{author}{\bibfnamefont{R.}~\bibnamefont{Gatto}},
  \bibinfo{author}{\bibfnamefont{M.}~\bibnamefont{Mannarelli}},
  \bibnamefont{and} \bibinfo{author}{\bibfnamefont{G.}~\bibnamefont{Nardulli}},
  \bibinfo{journal}{Phys. Rev.} \textbf{\bibinfo{volume}{D66}},
  \bibinfo{pages}{014006} (\bibinfo{year}{2002}{\natexlab{a}}),
  \eprint{hep-ph/0201059}.

\bibitem[{\citenamefont{Casalbuoni
  et~al.}(2002{\natexlab{b}})\citenamefont{Casalbuoni, Gatto, and
  Nardulli}}]{Casalbuoni:2002hr}
\bibinfo{author}{\bibfnamefont{R.}~\bibnamefont{Casalbuoni}},
  \bibinfo{author}{\bibfnamefont{R.}~\bibnamefont{Gatto}}, \bibnamefont{and}
  \bibinfo{author}{\bibfnamefont{G.}~\bibnamefont{Nardulli}},
  \bibinfo{journal}{Phys. Lett.} \textbf{\bibinfo{volume}{B543}},
  \bibinfo{pages}{139} (\bibinfo{year}{2002}{\natexlab{b}}),
  \eprint{hep-ph/0205219}.

\bibitem[{\citenamefont{Casalbuoni
  et~al.}(2002{\natexlab{c}})\citenamefont{Casalbuoni, Fabiano, Gatto,
  Mannarelli, and Nardulli}}]{Casalbuoni:2002my}
\bibinfo{author}{\bibfnamefont{R.}~\bibnamefont{Casalbuoni}},
  \bibinfo{author}{\bibfnamefont{E.}~\bibnamefont{Fabiano}},
  \bibinfo{author}{\bibfnamefont{R.}~\bibnamefont{Gatto}},
  \bibinfo{author}{\bibfnamefont{M.}~\bibnamefont{Mannarelli}},
  \bibnamefont{and} \bibinfo{author}{\bibfnamefont{G.}~\bibnamefont{Nardulli}},
  \bibinfo{journal}{Phys. Rev.} \textbf{\bibinfo{volume}{D66}},
  \bibinfo{pages}{094006} (\bibinfo{year}{2002}{\natexlab{c}}),
  \eprint{hep-ph/0208121}.

\bibitem[{\citenamefont{Combescot and Mora}(2003)}]{combescot:2003wz}
\bibinfo{author}{\bibfnamefont{R.}~\bibnamefont{Combescot}} \bibnamefont{and}
  \bibinfo{author}{\bibfnamefont{C.}~\bibnamefont{Mora}}
  (\bibinfo{year}{2003}), \eprint{cond-mat/0311042}.

\bibitem[{\citenamefont{Hong}(2000{\natexlab{a}})}]{Hong:1998tn}
\bibinfo{author}{\bibfnamefont{D.~K.} \bibnamefont{Hong}},
  \bibinfo{journal}{Phys. Lett.} \textbf{\bibinfo{volume}{B473}},
  \bibinfo{pages}{118} (\bibinfo{year}{2000}{\natexlab{a}}),
  \eprint{hep-ph/9812510}.

\bibitem[{\citenamefont{Hong}(2000{\natexlab{b}})}]{Hong:1999ru}
\bibinfo{author}{\bibfnamefont{D.~K.} \bibnamefont{Hong}},
  \bibinfo{journal}{Nucl. Phys.} \textbf{\bibinfo{volume}{B582}},
  \bibinfo{pages}{451} (\bibinfo{year}{2000}{\natexlab{b}}),
  \eprint{hep-ph/9905523}.

\bibitem[{\citenamefont{Beane et~al.}(2000)\citenamefont{Beane, Bedaque, and
  Savage}}]{Beane:2000ms}
\bibinfo{author}{\bibfnamefont{S.~R.} \bibnamefont{Beane}},
  \bibinfo{author}{\bibfnamefont{P.~F.} \bibnamefont{Bedaque}},
  \bibnamefont{and} \bibinfo{author}{\bibfnamefont{M.~J.}
  \bibnamefont{Savage}}, \bibinfo{journal}{Phys. Lett.}
  \textbf{\bibinfo{volume}{B483}}, \bibinfo{pages}{131} (\bibinfo{year}{2000}),
  \eprint{hep-ph/0002209}.

\bibitem[{\citenamefont{Casalbuoni
  et~al.}(2001{\natexlab{b}})\citenamefont{Casalbuoni, Gatto, and
  Nardulli}}]{Casalbuoni:2000na}
\bibinfo{author}{\bibfnamefont{R.}~\bibnamefont{Casalbuoni}},
  \bibinfo{author}{\bibfnamefont{R.}~\bibnamefont{Gatto}}, \bibnamefont{and}
  \bibinfo{author}{\bibfnamefont{G.}~\bibnamefont{Nardulli}},
  \bibinfo{journal}{Phys. Lett.} \textbf{\bibinfo{volume}{B498}},
  \bibinfo{pages}{179} (\bibinfo{year}{2001}{\natexlab{b}}),
  \eprint{hep-ph/0010321}.

\bibitem[{\citenamefont{Schafer}(2003{\natexlab{a}})}]{Schafer:2003jn}
\bibinfo{author}{\bibfnamefont{T.}~\bibnamefont{Schafer}},
  \bibinfo{journal}{Nucl. Phys.} \textbf{\bibinfo{volume}{A728}},
  \bibinfo{pages}{251} (\bibinfo{year}{2003}{\natexlab{a}}),
  \eprint{hep-ph/0307074}.

\bibitem[{\citenamefont{Casalbuoni}(2001)}]{Casalbuoni:2001dw}
\bibinfo{author}{\bibfnamefont{R.}~\bibnamefont{Casalbuoni}},
  \bibinfo{journal}{AIP Conf. Proc.} \textbf{\bibinfo{volume}{602}},
  \bibinfo{pages}{358} (\bibinfo{year}{2001}), \eprint{hep-th/0108195}.

\bibitem[{\citenamefont{Schafer}(2003{\natexlab{b}})}]{Schafer:2003yh}
\bibinfo{author}{\bibfnamefont{T.}~\bibnamefont{Schafer}},
  \bibinfo{journal}{ECONF} \textbf{\bibinfo{volume}{C030614}},
  \bibinfo{pages}{038} (\bibinfo{year}{2003}{\natexlab{b}}),
  \eprint{hep-ph/0310176}.

\bibitem[{\citenamefont{Casalbuoni et~al.}(2003)}]{Casalbuoni:2003sa}
\bibinfo{author}{\bibfnamefont{R.}~\bibnamefont{Casalbuoni}}
  \bibnamefont{et~al.}, \bibinfo{journal}{Phys. Lett.}
  \textbf{\bibinfo{volume}{B575}}, \bibinfo{pages}{181} (\bibinfo{year}{2003}),
  \eprint{hep-ph/0307335}.

\end{thebibliography}

\end{document}